\documentclass[final,3p,times]{elsarticle}

\usepackage{amssymb}

\usepackage{amssymb}
\usepackage[]{amsmath}
\usepackage{graphics}
\usepackage{amssymb}
\usepackage[]{amsmath}
\usepackage{amsthm}
\usepackage{setspace}
\usepackage{epsfig}
\usepackage{subfigure}
\usepackage{cases}
\usepackage{graphicx}

\biboptions{numbers,sort&compress}

\usepackage{amsmath}
\usepackage{times}
\usepackage{anysize}
\marginsize{2.5cm}{2.5cm}{1cm}{2cm}

\selectfont

\begin{document}

\begin{frontmatter}



\title{Nonlocal symmetry, Darboux transformation and soliton-cnoidal wave interaction solution for the shallow water wave equation }

\author{ Junchao Chen \fnref{label11} }
\ead{junchaochen@aliyun.com}
\author{ Zhengyi Ma \fnref{label11} }
\author{ Yahong Hu \fnref{label11} }
%
%
%
%
\cortext[cor1]{Corresponding author.}
\address[label11]{Department of Mathematics, Lishui University, Lishui, 323000, People¡¯s Republic of China}

\begin{abstract}
In classical shallow water wave (SWW) theory, there exist two integrable one-dimensional SWW equation [Hirota-Satsuma (HS) type and Ablowitz-Kaup-Newell-Segur (AKNS) type] in the Boussinesq approximation.
In this paper, we mainly focus on the integrable SWW equation of AKNS type.
The nonlocal symmetry in form of square spectral function
is derived starting from its Lax pair.
Infinitely many nonlocal symmetries are presented by introducing the arbitrary spectrum parameter.
These nonlocal symmetries can be localized and the SWW equation is extended to enlarged system with auxiliary dependent variables.
Then Darboux transformation for the prolonged system is found by solving the initial value problem.
Similarity reductions related to the nonlocal symmetry and explicit group invariant solutions are obtained.
It is shown that the soliton-cnoidal wave interaction solution, which represents soliton lying on a cnoidal periodic wave background,
can be obtained analytically.
Interesting characteristics of the interaction solution between soliton and cnoidal periodic wave are displayed graphically.

\end{abstract}

\end{frontmatter}


\section{Introduction}

The shallow water wave (SWW) theory is devoted to study the surface wave in shallow water
where the wavelength of the water is much larger than the local water depth.
Then the SWWs are termed as the flow at the free surface of a body of shallow water
under the force of gravity, or to the flow below a horizontal pressure surface in a fluid \cite{hereman2012shallow,craik2004origins,stokes1847theory}.
As one of the pioneers of hydrodynamics, Stokes \cite{stokes1847theory} derived the equations for the motion
of incompressible, inviscid fluid, subject to a constant vertical
gravitational force, where the fluid is bounded below
by an impermeable bottom and above by a free surface.
Starting from these fundamental equations and by making
further simplifying assumptions, various shallow water
wave models, especially some completely integrable nonlinear partial differential equations, can be derived at different levels of
approximation.
Such shallow water models
are widely used in oceanography and atmospheric science.
The classical reduced cases mainly include the Korteweg-de Vries(KdV) equation \cite{korteweg1895change}, the regularized long-wave equation(or the Benjamin-Bona-Mahony equation) \cite{benjamin1972model}, the one-dimensional (1D) SWW equation \cite{ablowitz1974inverse,hirota1976soliton} and the Camassa-Holm equation \cite{camassa1993integrable}.
Among these nonlinear equations, the 1D SWW equation
\begin{eqnarray}
\label{sww-non-02} U_{xxt}+\alpha UU_t + \beta U_x \partial^{-1}_xU_t-U_t-U_x=0,
\end{eqnarray}
with $\alpha$ and $\beta$ being arbitrary nonzero-constants,
arises from the Boussinesq approximation in which vertical variations
of the static density are neglected.
The integral term in Eq. (\ref{sww-non-02}) can be removed by introducing the potential
$u$. By taking $u_x=U$, the nonlocal form (\ref{sww-non-02}) can be written as
\begin{eqnarray}
\label{sww-non-01} u_{xxxt} +\alpha u_x u_{xt}+ \beta u_t u_{xx} -u_{xt}-u_{xx}=0,
\end{eqnarray}
There are two special cases of this equation which are completely integrable:
(i) when $\alpha=2\beta$, Eq. (\ref{sww-non-01}) becomes
\begin{eqnarray}
\label{sww-non-03} u_{xxxt} +2\beta u_x u_{xt}+ \beta u_t u_{xx} -u_{xt}-u_{xx}=0,
\end{eqnarray}
which is called as the SWW-Ablowitz-Kaup-Newell-Segur (SWW-AKNS) equation \cite{ablowitz1974inverse}, since
Ablowitz, Kaup, Newell and Segur showed that Eq.(\ref{sww-non-03}) is solvable via inverse scattering method;
(ii) when $\alpha=\beta$, Eq.(\ref{sww-non-01}) reduces to
\begin{eqnarray}
\label{sww-non-03-hs} u_{xxxt} +\beta u_x u_{xt}+ \beta u_t u_{xx} -u_{xt}-u_{xx}=0,
\end{eqnarray}
which is referred to the SWW-Hirota-Satsuma (SWW-HS) equation \cite{hirota1976soliton}, because Hirota and
Satsuma studied Eq.(\ref{sww-non-03-hs}) by using Hirota's bilinear method.
There is no scaling transformation that reduces (\ref{sww-non-03}) to (\ref{sww-non-03-hs}) so that the SWW-AKNS and SWW-HS equations are fundamentally different.
The SWW Eq.(\ref{sww-non-01}) was also discussed by Hietarinta\cite{hietarinta1990partially} in the Hirota's bilinear form.
Clarkson and Mansfield \cite{clarkson1994shallow,clarkson1995symmetry} have carried out classical and non-classical symmetry reductions
for the SWW Eq.(\ref{sww-non-01}) and further obtained exact solutions with the rich variety of qualitative behaviours.

The symmetry method is one of most powerful tools in differential equations.
For a given differential system,
the general symmetry groups and associated reductions can be realized by using the classical or nonclassical Lie group method \cite{olver1993applications,bluman2010applications}.
Apart from these local symmetries, the study of nonlocal symmetry have received much attention
and some effective techniques to find nonlocal symmetries have been developed\cite{krasilshchik1989nonlocal}.
In particular, for the integrable system, there exists a class of nonlocal symmetries related to the recursion operator and its inverse, Lax pair, B\"{a}cklund transformation, the conformal invariant form and pseudopotentials \cite{guthrie1994recursion,galas1992new,gorka2011modified,hu2009nolocal,lou1997nolocal,lou2009conformal}.
When such nonlocal symmetries are converted into local ones, the application of symmetry group is further extended \cite{lou2012nonlocal,hu2012explicit,gao2013bosonization,cheng2014interactions,chen2014nonlocal,chen2014nonlocal1,cheng2014interactions-pre,tang2015nonlocal,wang2014oblique,cheng2015nolocal,li2014darboux}.
For instance, the second type of Darboux transformation for the KdV equation was found by localizing the nonlocal symmetry related to the spectral function \cite{hu2012explicit,cheng2014interactions,li2014darboux}.
Therefore, the first objective of our paper is to derive nonlocal symmetry for integrable SWW-AKNS equation (\ref{sww-non-03}) from its Lax pair, and further to localize such nonlocal symmetry by introducing the potential variables.
Then we start from local symmetries to construct the corresponding Darboux transformation which, to the best our knowledge, was never reported before.

Solitary waves and cnoidal periodic waves are two typical continuous waves widely appearing in many physical fields, especially in nonlinear integrable system.
Considering the potential applicability of such two waves, it is desirable to
have one kind of more generalized solution to describe the real complicated waves.
One possible case is to construct an interaction solution between solitons and cnoidal periodic waves \cite{shin2005dark}.
However, it is not easy to obtain this interaction solution via other traditional methods such as Darboux transformation, B\"{a}cklund and Hirota's bilinear method.
Very recently, starting from symmetry reductions of nonlocal symmetry,
various exact interaction solutions among solitons and other complicated
waves, especially, the soliton-cnoidal periodic wave interaction solution have been found from the group invariant solutions \cite{lou2012nonlocal,hu2012explicit,gao2013bosonization,cheng2014interactions,chen2014nonlocal,chen2014nonlocal1,cheng2014interactions-pre,tang2015nonlocal,wang2014oblique,cheng2015nolocal}.
For the SWW-AKNS equation, in the localizing procedure of the nonlocal symmetry,
the original equation is extended to a prolonged system with four dependent variables.
After the localization of the nonlocal symmetry, we devote to construct
interacting solution between soliton and cnoidal periodic wave from the associated symmetry reductions from the classical Lie point symmetry of the enlarged system.

The paper is arranged as follows.
In Section 2, we first derive the nonlocal symmetry for the SWW-AKNS equation from it Lax pair.
Then the nonlocal symmetry is localized by extending the original equation to an enlarged system and
the finite symmetry transformation is obtain.
Furthermore, due to the existence of the spectrum parameter, infinitely many nonlocal symmetries are given and
the corresponding finite symmetry transformation, i.e., Darboux transformation of the second type is provided.
For the enlarged system, the general Lie point symmetry is performed and the corresponding similarity reductions are followed in Section 3.
In section 4,
explicit invariant solutions regarding the interactions between soltion and cnoidal periodic waves are
presented and displayed graphically.
In the last section, some conclusions and discussions are given.

\section{Nonlocal symmetry, localization and Darboux transformation}

For the SWW-AKNS Eq.(\ref{sww-non-03}),  its Lax pair reads
\begin{eqnarray}
\label{sww-non-04} && \psi_{xx} + \frac{1}{2}\beta u_x\psi-\lambda\psi=0,\\
\label{sww-non-05} && (4\lambda-1)\psi_t - (1-\beta u_t)\psi_x - \frac{1}{2}\beta u_{xt}\psi=0,
\end{eqnarray}
where $\psi$ is the spectral function and $\lambda$ is the spectral parameter.

A symmetry of Eq.(\ref{sww-non-03}) is defined as a solution of the linearized equation
\begin{eqnarray}
\label{sww-non-06} \sigma_{xxxt}+2\beta (u_x\sigma_x)_t + \beta u_t \sigma_{xx} +\beta \sigma_t u_{xx} - \sigma_{xx}-\sigma_{xt}=0.
\end{eqnarray}
This means that Eq.(\ref{sww-non-03}) is form invariant under the transformation $u\rightarrow u+\epsilon \sigma$ with the infinitesimal parameter.

By the direct calculation, we find that Eq.(\ref{sww-non-03}) has the following square spectral function symmetry
\begin{eqnarray}
\label{sww-non-07} \sigma=\psi^2.
\end{eqnarray}

In order to localize the nonlocal symmetry (\ref{sww-non-07}),
we need to introduce the auxiliary variables $p=p(x,t)$ and $\phi=\phi(x,t)$ with the relations
\begin{eqnarray}
\label{sww-non-08} && \phi=\psi_x,\ \  p_x=\psi^2.
\end{eqnarray}
Then, we list the linearized equations of the prolonged system (\ref{sww-non-03}), (\ref{sww-non-04}) and (\ref{sww-non-08}) as follows
\begin{eqnarray}
\label{sww-non-09} && \sigma^u_{xxxt}+2\beta (u_x\sigma^u_x)_t + \beta u_t \sigma^u_{xx} +\beta \sigma^u_t u_{xx} - \sigma^u_{xx}-\sigma^u_{xt}=0,\\
\label{sww-non-10} && \sigma^\psi_{xx}+\frac{1}{2}\beta (\sigma^u_x \psi + u_x \sigma^\psi)-\lambda\sigma^\psi=0,\\
\label{sww-non-11} &&  \sigma^\psi_x-\sigma^\phi=0,\\
\label{sww-non-12} && \sigma^p_x-2\sigma^\psi \psi=0.
\end{eqnarray}
It is easy to verify that Eqs.(\ref{sww-non-09})--(\ref{sww-non-12}) have a solution
\begin{eqnarray}
\label{sww-non-13} \sigma=\left( \begin{array}{c}
                \sigma^u \\
                \sigma^\psi \\
                \sigma^\phi \\
                \sigma^p
              \end{array}
 \right)
 =\left( \begin{array}{c}
                \psi^2 \\
                -\frac{\beta}{4}p\psi \\
                -\frac{\beta}{4}(\psi^3+p\phi) \\
                -\frac{\beta}{4}p^2
              \end{array}
 \right),
\end{eqnarray}
Thus the nonlocal symmetry (\ref{sww-non-07}) becomes the local Lie point symmetry (\ref{sww-non-13}) of the prolonged system (\ref{sww-non-03}), (\ref{sww-non-04}) and (\ref{sww-non-08}).

Here, we prefer to mention another important fact.
From (\ref{sww-non-04}), (\ref{sww-non-05}) and (\ref{sww-non-08}),
one can find that the introduced variable $p$ just satisfies the Schwartzian form of the SWW-AKNS equation
\begin{eqnarray*}
\left[ \frac{p_{xxx}}{p_x}-\frac{3}{2}\left(\frac{p_{xx}}{p_x}\right)^2 \right]_t+(4\lambda-1)\left(\frac{p_t}{p_x} \right)_x=0,
\end{eqnarray*}
which possesses the M\"{o}bious (conformal) invariance property.
Thus $\sigma^p= -\frac{\beta}{4}p^2$ in (\ref{sww-non-13}) is the infinitesimal form of the M\"{o}bious transformation.

By solving the initial value problem
\begin{eqnarray}\label{sww-non-14}
\nonumber &&\frac{dU(\epsilon)}{d\epsilon}=\Psi(\epsilon)^2,\ \ U(0)=u,\\
&&\frac{d\Psi(\epsilon)}{d\epsilon}=-\frac{\beta}{4}P(\epsilon)\Psi(\epsilon),\ \ \Psi(0)=\psi,\\
\nonumber &&\frac{d\Phi(\epsilon)}{d\epsilon}=-\frac{\beta}{4}[\Psi(\epsilon)^3+P(\epsilon)\Phi(\epsilon)],\ \ \Phi(0)=\phi,\\
\nonumber &&\frac{dP(\epsilon)}{d\epsilon}=-\frac{\beta}{4}P(\epsilon)^2,\ \ P(0)=p,
\end{eqnarray}
one can derive the finite transformation straightforwardly, which can be stated in the
following theorem.

\textbf{Theorem 1} If $\{u,\psi,\phi,p\}$ is a solution of the extended system (\ref{sww-non-03}), (\ref{sww-non-04}) and (\ref{sww-non-08}),
so is $\{U(\epsilon),\Psi(\epsilon)$,$\Phi(\epsilon)$,$P(\epsilon)\}$ with
\begin{eqnarray}
\label{sww-non-15}&&U(\epsilon)=u+\frac{4\epsilon \psi^2}{\epsilon \beta p+4},\\
\label{sww-non-16}&&\Psi(\epsilon)=\frac{4\psi}{\epsilon \beta p+4},\\
\label{sww-non-17}&&\Phi(\epsilon)=\frac{4\phi}{\epsilon \beta p+4}-\frac{4\epsilon\beta\psi^3}{(\epsilon \beta p+4)^2},\\
\label{sww-non-18}&&P(\epsilon)=\frac{4p}{\epsilon \beta p+4},
\end{eqnarray}
where $\epsilon$ is an arbitrary group parameter.

Due to the arbitrariness of the spectral parameter $\lambda$ in the nonlocal symmetry (\ref{sww-non-07}),
we can obtain infinitely many nonlocal symmetry
\begin{eqnarray}
\label{sww-non-19}\sigma^u_n=\sum^n_{i=1}c_i\psi^2_i,\ \ n=1,2,\cdots,
\end{eqnarray}
where $\psi_i$, $i=1,2,\cdots,n$ are spectral functions in Lax pairs (\ref{sww-non-04})-(\ref{sww-non-05})
with different spectral parameters $\lambda_i\neq\lambda_j$ for $\forall j\neq i$.
Further, we introduce the following enlarged system
\begin{eqnarray}
\label{sww-non-20}&& u_{xxxt} +2\beta u_x u_{xt}+ \beta u_t u_{xx} -u_{xt}-u_{xx}=0,\\
\label{sww-non-21}&& \psi_{i,xx} + \frac{1}{2}\beta u_x\psi_i-\lambda_i\psi_i=0,\\
\label{sww-non-22}&& (4\lambda_i-1)\psi_{i,t} - (1-\beta u_t)\psi_{i,x} - \frac{1}{2}\beta u_{xt}\psi_i=0,\\
\label{sww-non-23}&& \phi_i=\psi_{i,x},\ \  p_{i,x}=\psi_i^2,
\end{eqnarray}
for $i=1,2,\cdots,n$.
Then the nonlocal symmetry (\ref{sww-non-19}) becomes local Lie point symmetries
\begin{eqnarray}\label{sww-non-24}
\nonumber &&\sigma^u=\sum^n_{i=1}c_i\psi^2_i,\\
&&\sigma^{\psi_j}=-\frac{\beta}{4}c_j\psi_jp_j+\frac{\beta}{4}\sum^n_{i\neq j} \frac{c_i\psi_i(\psi_i\phi_j-\psi_j\phi_i)}{\lambda_i-\lambda_j},\\
\nonumber &&\sigma^{\phi_j}=-\frac{\beta}{4}c_j(\psi^3_j+p_j\phi_j)-\frac{\beta}{4}\sum^n_{i\neq j}c_i[\psi^2_i\psi_j-\frac{\phi_i(\psi_i\phi_j-\psi_j\phi_i)}{\lambda_i-\lambda_j}],\\
\nonumber &&\sigma^{p_j}=-\frac{\beta}{4}c_j p^2_j -\frac{\beta}{4}\sum^n_{i\neq j}\frac{c_i(\psi_i\phi_j-\psi_j\phi_i)^2}{(\lambda_i-\lambda_j)^2},
\end{eqnarray}
with $j=1,2,\cdots,n$, which is a solution of linearized system
\begin{eqnarray}
\label{sww-non-25}&& \sigma^u_{xxxt}+2\beta (u_x\sigma^u_x)_t + \beta u_t \sigma^u_{xx} +\beta \sigma^u_t u_{xx} - \sigma^u_{xx}-\sigma^u_{xt}=0,\\
\label{sww-non-26}&& \sigma^{\psi_j}_{xx}+\frac{1}{2}\beta (\sigma^u_x \psi_j + u_x \sigma^{\psi_j})-\lambda_j\sigma^{\psi_j}=0,\\
\label{sww-non-27}&&  \sigma^{\psi_j}_x-\sigma^{\phi_j}=0,\\
\label{sww-non-28}&& \sigma^{p_j}_x-2\sigma^{\psi_j} \psi_j=0,
\end{eqnarray}
for $j=1,2,\cdots,n$.
Here the local symmetries (\ref{sww-non-24}) can be verified by the direct calculation.
Without loss of generality, we fix $c_i\neq0$ and $c_k=0 (k\neq i)$,
then
\begin{eqnarray}
\label{sww-non-29}\sigma^u=c_i\psi^2_i,
\end{eqnarray}
which is a known solution of (\ref{sww-non-25}).
For $i=j$, substituting (\ref{sww-non-29}) into (\ref{sww-non-26}) yields
\begin{eqnarray}
\label{sww-non-30}\sigma^{\psi_j}_{xx}+\frac{1}{2}\beta (c_j\psi_j (\psi^2_j)_x  + u_x \sigma^{\psi_j})-\lambda_j\sigma^{\psi_j}=0.
\end{eqnarray}
By eliminating $u_x$ in (\ref{sww-non-30}) through (\ref{sww-non-21}) with $i=j$, one has
\begin{eqnarray}
\label{sww-non-31}\sigma^{\psi_j}_{xx}+\frac{1}{2}\beta c_j\psi_j (\psi^2_j)_x - \frac{\psi_{j,xx}}{\psi_j}\sigma^{\psi_j}=0.
\end{eqnarray}
It is easy to check that Eqs.(\ref{sww-non-26})--(\ref{sww-non-28}) have the solution
\begin{eqnarray}
\label{sww-non-32}\sigma^{\psi_j}=-\frac{\beta}{4}c_j\psi_jp_j,\ \ \sigma^{\phi_j}=-\frac{\beta}{4}c_j(\psi^3_j+p_j\phi_j),\ \ \sigma^{p_j}=-\frac{\beta}{4}c_j p^2_j.
\end{eqnarray}
For $i\neq j$, we have
\begin{eqnarray}
\label{sww-non-33} \sigma^{\psi_i}_{xx}+\frac{1}{2}\beta c_j\psi_i (\psi^2_j)_x  -(\frac{\psi_{j,xx}}{\psi_j}-\lambda_j+\lambda_i) \sigma^{\psi_i}=0.
\end{eqnarray}
By using the relation that comes from the Lax pairs
\begin{eqnarray}
\label{sww-non-34}\frac{\psi_{i,xx}}{\psi_i}-\lambda_i=\frac{\psi_{j,xx}}{\psi_j}-\lambda_j,
\end{eqnarray}
one can easily obtain
\begin{eqnarray}\label{sww-non-35}
\sigma^{\psi_i}=\frac{\beta}{4} \frac{c_i\psi_i(\psi_i\phi_j-\psi_j\phi_i)}{\lambda_i-\lambda_j},\ \
\sigma^{\phi_i}=-\frac{\beta}{4}c_i[\psi^2_i\psi_j-\frac{\phi_i(\psi_i\phi_j-\psi_j\phi_i)}{\lambda_i-\lambda_j}],\ \
\sigma^{p_i}= -\frac{\beta}{4}\frac{c_i(\psi_i\phi_j-\psi_j\phi_i)^2}{(\lambda_i-\lambda_j)^2}.
\end{eqnarray}

Now, we can consider the following initial value problem
\begin{eqnarray}\label{sww-non-36}
\nonumber &&\frac{dU(\epsilon)}{d\epsilon}=\sum^n_{i=1}c_i\Psi^2_i(\epsilon),\\
\nonumber &&\frac{d\Psi_j(\epsilon)}{d\epsilon}=-\frac{\beta}{4}c_j\Psi_j(\epsilon)P_j(\epsilon)+\frac{\beta}{4}\sum^n_{i\neq j} \frac{c_i\Psi_i(\epsilon)[\Psi_i(\epsilon)\Phi_j(\epsilon)-\Psi_j(\epsilon)\Phi_i(\epsilon)]}{\lambda_i-\lambda_j},\\
&&\frac{d\Phi_j(\epsilon)}{d\epsilon}=-\frac{\beta}{4}c_j[\Psi^3_j(\epsilon){+} P_j(\epsilon)\Phi_j(\epsilon)]-\frac{\beta}{4}\sum^n_{i\neq j}c_i\{\Psi^2_i(\epsilon)\Psi_j(\epsilon)-\frac{\Phi_i(\epsilon)[\Psi_i(\epsilon)\Phi_j(\epsilon)-\Psi_j(\epsilon)\Phi_i(\epsilon)]}{\lambda_i-\lambda_j}\},\ \ \ \\
\nonumber &&\frac{d P_j(\epsilon)}{d\epsilon}=-\frac{\beta}{4}c_j P^2_j(\epsilon) -\frac{\beta}{4}\sum^n_{i\neq j}\frac{c_i[\Psi_i(\epsilon)\Phi_j(\epsilon)-\Psi_j(\epsilon)\Phi_i(\epsilon)]^2}{(\lambda_i-\lambda_j)^2},\\
\nonumber && U(0)=u,\ \ \Psi_j(0)=\psi_j,\ \ \Phi_j(0)=\phi_j,\ \ P_j(0)=p_j,
\end{eqnarray}
for $j=1,\cdots,n$.
By solving Eqs.(\ref{sww-non-36}), one can obtain the corresponding finite transformation, which is summarized to the theorem as follows.

\textbf{Theorem 2}
If $\{u,\psi_i,\phi_i,p_i,\ \ i=1,2,\cdots,n \}$ is a solution of the enlarged system (\ref{sww-non-20})--(\ref{sww-non-23}),
so is \\
$\{ U(\epsilon)$,$\Psi_i(\epsilon)$,$\Phi_i(\epsilon),$$P_i(\epsilon),$$ i=1,2,\cdots,n \}$ with
\begin{eqnarray}
\label{sww-non-37} && U(\epsilon)=u+\frac{4}{\beta}(\ln|M|)_x,\\
\label{sww-non-38} && \Psi_i(\epsilon)=-4\frac{|N_i|}{|M|},\\
\label{sww-non-39} && \Phi_i(\epsilon)=\Psi_{i,x}(\epsilon),\\
\label{sww-non-40} && P_i(\epsilon)=4\frac{|M_i|}{|M|},
\end{eqnarray}
where $|M|,|M_i|$ and $|N_i|$ are determinants of the matrices defined by
\begin{eqnarray*}
&&\hspace{-0.8cm} M=\left(
       \begin{array}{cccccc}
         c_1\epsilon\beta p_1+4 & c_1\epsilon \beta w_{12} & \cdots & c_1\epsilon \beta w_{1j} & \cdots & c_1\epsilon \beta w_{1n} \\
         c_2\epsilon \beta w_{12} & c_2\epsilon\beta p_2+4 & \cdots & c_2\epsilon \beta w_{2j} & \cdots & c_2\epsilon \beta w_{2n} \\
         \vdots & \vdots & \vdots & \vdots & \vdots & \vdots \\
         c_j\epsilon \beta w_{1j} & c_j\epsilon \beta w_{2j} & \cdots & c_j\epsilon\beta p_j+4 & \cdots & c_j\epsilon \beta w_{jn} \\
         \vdots & \vdots & \vdots & \vdots & \vdots & \vdots \\
         c_n\epsilon \beta w_{1n} &  c_n\epsilon \beta w_{2n} & \cdots &  c_n\epsilon \beta w_{jn} & \cdots & c_n\epsilon\beta p_n+4 \\
       \end{array}
     \right),
\end{eqnarray*}
\begin{eqnarray*}
&&\hspace{-0.8cm} M_i=\left(
       \begin{array}{cccccccc}
         c_1\epsilon\beta p_1+4 & c_1\epsilon \beta w_{12} & \cdots & c_1\epsilon \beta w_{1,i-1} & c_1\epsilon \beta w_{1i} & c_1\epsilon \beta w_{1,i+1} & \cdots & c_1\epsilon \beta w_{1n} \\
         c_2\epsilon \beta w_{12} & c_2\epsilon\beta p_2+4 & \cdots & c_2\epsilon \beta w_{2,i-1} & c_2\epsilon \beta w_{2i} & c_2\epsilon \beta w_{2,i+1} & \cdots & c_2\epsilon \beta w_{2n} \\
         \vdots & \vdots & \vdots & \vdots & \vdots & \vdots & \vdots & \vdots \\
         c_{i-1}\epsilon \beta w_{1,i-1} & c_{i-1}\epsilon \beta w_{2,i-1} & \cdots & c_{i-1}\epsilon\beta p_{i-1}+4 & c_{i-1}\epsilon \beta w_{i-1,i} & c_{i-1}\epsilon \beta w_{i-1,i+1} & \cdots & c_{i-1}\epsilon \beta w_{i-1,n} \\
         w_{1i} & w_{2i} & \cdots & w_{i,i-1} & p_i & w_{i,i+1} & \cdots & w_{in}\\
         c_{i+1}\epsilon \beta w_{1,i+1} & c_{i+1}\epsilon \beta w_{2,i+1} & \cdots &  c_{i+1}\epsilon \beta w_{i-1,i+1} & c_{i+1}\epsilon \beta w_{i,i+1} & c_{i+1}\epsilon\beta p_{i+1}+4 &  \cdots & c_{i+1}\epsilon \beta w_{i+1,n} \\
         \vdots & \vdots & \vdots & \vdots & \vdots & \vdots & \vdots & \vdots \\
         c_n\epsilon \beta w_{1n} &  c_n\epsilon \beta w_{2n} & \cdots & c_n\epsilon \beta w_{i-1,n} &  c_n\epsilon \beta w_{in} & c_n\epsilon \beta w_{i+1,n} & \cdots & c_n\epsilon\beta p_n+4 \\
       \end{array}
     \right),
\end{eqnarray*}
\begin{eqnarray*}
&&\hspace{-0.8cm} N_i=\left(
       \begin{array}{cccccccc}
         c_1\epsilon\beta p_1+4 & c_1\epsilon \beta w_{12} & \cdots & c_1\epsilon \beta w_{1,i-1} & c_1\epsilon \beta w_{1i} & c_1\epsilon \beta w_{1,i+1} & \cdots & c_1\epsilon \beta w_{1n} \\
         c_2\epsilon \beta w_{12} & c_2\epsilon\beta p_2+4 & \cdots & c_2\epsilon \beta w_{2,i-1} & c_2\epsilon \beta w_{2i} & c_2\epsilon \beta w_{2,i+1} & \cdots & c_2\epsilon \beta w_{2n} \\
         \vdots & \vdots & \vdots & \vdots & \vdots & \vdots & \vdots & \vdots \\
         c_{i-1}\epsilon \beta w_{1,i-1} & c_{i-1}\epsilon \beta w_{2,i-1} & \cdots & c_{i-1}\epsilon\beta p_{i-1}+4 & c_{i-1}\epsilon \beta w_{i-1,i} & c_{i-1}\epsilon \beta w_{i-1,i+1} & \cdots & c_{i-1}\epsilon \beta w_{i-1,n} \\
         \psi_1 & \psi_2 & \cdots & \psi_{i-1} & -\psi_i & \psi_{i+1} & \cdots & \psi_{n}\\
         c_{i+1}\epsilon \beta w_{1,i+1} & c_{i+1}\epsilon \beta w_{2,i+1} & \cdots &  c_{i+1}\epsilon \beta w_{i-1,i+1} & c_{i+1}\epsilon \beta w_{i,i+1} & c_{i+1}\epsilon\beta p_{i+1}+4 &  \cdots & c_{i+1}\epsilon \beta w_{i+1,n} \\
         \vdots & \vdots & \vdots & \vdots & \vdots & \vdots & \vdots & \vdots \\
         c_n\epsilon \beta w_{1n} &  c_n\epsilon \beta w_{2n} & \cdots & c_n\epsilon \beta w_{i-1,n} &  c_n\epsilon \beta w_{in} & c_n\epsilon \beta w_{i+1,n} & \cdots & c_n\epsilon\beta p_n+4 \\
       \end{array}
     \right),
\end{eqnarray*}
and
\begin{eqnarray*}
w_{ij}=\frac{\psi_i\phi_j-\phi_i\psi_j}{\lambda_i-\lambda_j}.
\end{eqnarray*}

It should be point out that the finite transformation in the theorem is equivalent to
Darboux transformation of the second type (i.e., the binary DT or Levi transformation).
Thus, the multiple solitons can be obtained from Theorem 2.
For instance, we take the trivial solution $u=0$ of Eq.(\ref{sww-non-03}).
From (\ref{sww-non-21})-(\ref{sww-non-23}) with $\lambda_i=\frac{k_i}{4}$, one can find the special solution
\begin{eqnarray}\label{sww-non-41}
&& \psi_i=\exp[\frac{k_i}{2}x+\frac{k_it}{2(k^2_i-1)}],\ \
\phi_i=\frac{k_i}{2}\exp[\frac{k_i}{2}x+\frac{k_it}{2(k^2_i-1)}],\ \
p_i=\frac{1}{k_i}\exp(k_i x+\frac{k_it}{k^2_i-1}),
\end{eqnarray}
and then
\begin{eqnarray}\label{sww-non-42}
w_{ij}=-\frac{2}{k_i+k_j}\exp\left\{\frac{k_i+k_j}{2}[x+\frac{k_ik_j-1}{(k^2_i-1)(k^2_j-1)}t] \right\}.
\end{eqnarray}
Substituting (\ref{sww-non-41}) and (\ref{sww-non-42}) into the transformation (\ref{sww-non-37}), we can obtain the multi-soliton solution of SWW-AKNS equation.
For the case $n=2$, the determinant $|M|$ reads
\begin{eqnarray}\label{sww-non-43}
\nonumber |M|&=&16+4\epsilon\beta [ \frac{c_1}{k_1}\exp(k_1 x+\frac{k_1 t}{k^2_1-1}) + \frac{c_2}{k_2}\exp(k_2 x+\frac{k_2t}{k^2_2-1}) ]\\
&&+c_1c_2\beta\epsilon^2\frac{(k_1-k_2)^2}{k_1k_2(k_1+k_2)^2}\exp[(k_1+k_2)x+( \frac{k_1}{k^2_1-1}+\frac{k_2}{k^2_2-1})t ],
\end{eqnarray}
for the two-soliton solution.

\section{Similarity reductions with the nonlocal symmetry}


Symmetry reduction is one of the main purposes for calculating symmetries of differential equation(s).
For the SWW-AKNS equarion, its extended system has been listed in the last section.
It is easy to demonstrate that the general Lie point symmetry solution of
the extended SWW-AKNS system of Eqs. (\ref{sww-non-03}), (\ref{sww-non-04}), (\ref{sww-non-05}) and (\ref{sww-non-08}) has the
form

\begin{eqnarray}\label{sww-non-44}
&& \sigma=\left( \begin{array}{c}
                   \sigma^u \\
                   \sigma^\psi \\
                   \sigma^\phi \\
                   \sigma^p \\
                   \sigma^\lambda
                 \end{array}
 \right)
 =\left( \begin{array}{c}
                   (c_1x+c_2)u_x + f u_t -c_3 - \frac{1}{\beta}(f+c_1x+c_1t) +c_1u +c_4\psi^2 \\
                   (c_1x+c_2)\psi_x + f \psi_t -\frac{1}{4}(c_4\beta p+4c_5)\psi \\
                   (c_1x+c_2)\phi_x + f \phi_t -\frac{1}{4}(c_4\beta p -4c_1 +4c_5)\phi - \frac{1}{4}c_4\beta\psi^3 \\
                   (c_1x+c_2)p_x+ f p_t -c_6 - (c_1+2c_5)p - \frac{1}{4}c_4\beta p^2 \\
                   2c_1\lambda-\frac{1}{2}c_1
                 \end{array}
  \right),
\end{eqnarray}
where $f=f(t)$ is an arbitrary function of $t$.
Notice that one component of the symmetry (\ref{sww-non-44}) with $c_i = 0 (i = 1, 2, 3, 5,6)$, $f=0$ and
$c_4 = 1$ is just corresponding to the localized symmetry of the nonlocal symmetry (\ref{sww-non-13}).
Besides, it is found that the scaling of the augmented system must accompany the
transformation of the spectral parameter $\lambda$.
When $c_i = 0 (i = 3,4,5,6)$, the symmetry (\ref{sww-non-44}) is just the general Lie point symmetry of the SWW-AKNS Eq.(\ref{sww-non-03}) \cite{clarkson1994shallow,clarkson1995symmetry}.

To obtain group invariant solutions, one need to solve the invariant conditions $\sigma^{i}=0$ with $i=u,\psi,\phi,p$ and $\lambda$.
Here, we need to require $c_4\neq0$, which guarantees to give the group invariant solutions related to the nonlocal symmetry.
Meanwhile, from the condition $\sigma^{\lambda}=0$, namely, $2c_1\lambda-\frac{1}{2}c_1=0$, we have to choose $c_1=0$ since $\lambda$ cannot be equal to $\frac{1}{4}$.
It means that no scaling transformation happens for the extended SWW-AKNS system.
Without loss of generality, we redefine $k^2=4c^2_5-\beta c_4c_6$ and then there are two nontrivial symmetry reductions.

\textbf{Case 1} $k\neq0$. In this case, we set $f=1/\dot{g}$ and obtain the similarity solutions
\begin{eqnarray}
\label{sww-similar-solution11} && u=c_3g+\frac{t}{\beta}+U(z)-\frac{2c_4}{k}\Psi(z)^2\tanh\{\frac{k}{2}[g+P(z)]\},\\
&& \psi=\Psi(z){\rm sech}\{\frac{k}{2}[g+P(z)]\},\\
&& \phi=\Phi(z){\rm sech}\{\frac{k}{2}[g+P(z)]\}+\frac{\beta c_4}{2k}\Psi(z)^3{\rm sinh}\{\frac{k}{2}[g+P(z)]\}{\rm sech^2}\{\frac{k}{2}[g+P(z)]\},\\
\label{sww-similar-solution14}&& p=-\frac{2}{\beta c_4} \tanh\{\frac{k}{2}[g+P(z)]\} -\frac{4c_5}{\beta c_4},
\end{eqnarray}
with the similarity variable $z=x-c_2g$.
Substituting (\ref{sww-similar-solution11})-(\ref{sww-similar-solution14}) into the extended system leads to
\begin{eqnarray}
&& P_z(z)=Q(z),\ \
\Psi(z)=\pm k\sqrt{\frac{Q(z)}{\beta c_4}},\ \
\Phi(z)=\mp \frac{k Q_z(z)}{2\sqrt{-\beta c_4 Q(z)}},\\
&& U_z(z)=\frac{2\lambda}{\beta}-\frac{k^2Q(z)^2}{\beta}-\frac{Q_{zz}(z)}{\beta Q(z)}+\frac{Q^2_z(z)}{2\beta Q(z)^2},
\end{eqnarray}
where $Q(z)$ satisfies the ordinary differential equation
\begin{eqnarray}
\label{last-reduce-1} Q^2_z(z)=a_1Q(z)+a_2Q(z)^2+a_3Q(z)^3+a_4Q(z)^4,
\end{eqnarray}
with
\begin{eqnarray}
a_1=\frac{1-4\lambda}{c_2},\ \ a_2=12\lambda-2-\frac{2\beta c_3}{c_2},\ \ a_4=k^2.
\end{eqnarray}

\textbf{Case 2} $k=0$. In this case, we still set $f=1/\dot{g}$ and have the similarity solutions
\begin{eqnarray}
\label{sww-similar-solution21}&& u=c_3g+\frac{t}{\beta}+U(z)+\frac{c_4\Psi^2(z)}{g+P(z)},\\
&& \psi=\frac{\Psi(z)}{g+P(z)},\\
&& \phi=\frac{\Phi(z)}{g+P(z)}-\frac{\beta c_4 \Psi^3(z)}{4[g+P(z)]^2},\\
\label{sww-similar-solution24}&& p=-\frac{4}{\beta c_4 [g+P(z)]}  -\frac{4c_5}{\beta c_4},
\end{eqnarray}
with the similarity variable $z=x-c_2g$.
Substituting (\ref{sww-similar-solution21})-(\ref{sww-similar-solution24}) into the extended system yields
\begin{eqnarray}
&& P_z(z)=Q(z),\ \
\Psi(z)=\pm 2\sqrt{\frac{Q(z)}{\beta c_4}},\ \
\Phi(z)=\pm \frac{Q_z(z)}{\sqrt{\beta c_4 Q(z)}},\\
&& U_z(z)=\frac{2\lambda}{\beta}-\frac{Q_{zz}(z)}{\beta Q(z)}+\frac{Q^2_z(z)}{2\beta Q(z)^2},
\end{eqnarray}
where $Q(z)$ satisfies the ordinary differential equation
\begin{eqnarray}
\label{last-reduce-2} Q^2_z(z)=a_1Q(z)+a_2Q(z)^2+a_3Q(z)^3,
\end{eqnarray}
with
\begin{eqnarray*}
a_1=\frac{1-4\lambda}{c_2},\ \ a_2=12\lambda-2-\frac{2\beta c_3}{c_2}.
\end{eqnarray*}

\section{Explicit soltion-cnoidal periodic wave interaction solutions}

From the two reductions in the last section, we know that the explicit solution of the SWW-AKNS equation can be acquired by solving the last reduction equations (\ref{last-reduce-1}) and (\ref{last-reduce-2}), respectively.
The general solutions of Eqs.(\ref{last-reduce-1}) and (\ref{last-reduce-2}) are given in terms of Jacobi elliptic function.
Hence, two kinds of solutions represent the interactions among cnoidal periodic waves and soliton, rational waves.
To show more clearly of the interacting behavior occurring at the different types of nonlinear wave, we only provide two special solutions regarding the interactions between soltion and cnoidal periodic waves from the first reduction.

For the first case of symmetry reductions, it is known that the explicit solution of the SWW-AKNS equation (the original physical quantity) is finally expressed as
\begin{eqnarray}
\nonumber U&=&\frac{2\lambda}{\beta} + \frac{k^2Q(z)^2}{2\beta} - \frac{Q_{zz}(z)}{\beta Q(z)} + \frac{Q^2_z(z)}{2\beta Q(z)^2}
+\frac{2k}{\beta}Q_z(z)\tanh\left\{\frac{k}{2}\left[g+\int Q(z)dz \right]\right\} \\
&& -\frac{k^2}{\beta}Q^2(z)\tanh^2\left\{\frac{k}{2}\left[g+\int Q(z)dz \right]\right\},
\end{eqnarray}
where $Q(z)$ is determined by (\ref{last-reduce-1}) and $z=x-c_2g$.

\emph{Case 1} A simple solution of the reduction Eq. (\ref{last-reduce-1}) is given by
\begin{eqnarray}
&& Q(z)=b_0[1+m {\rm sn}(b_0 k z,m)],
\end{eqnarray}
with
\begin{eqnarray*}
&& a_3=-4b_0k^2,\ \ \beta=-\frac{c_2}{2c_3}\{b^2_0k^2[ 6b_0c_2 (m^2-1)-m^2+5 ]-1\},\ \
 \lambda=\frac{1}{4}-\frac{c_2}{2}b^3_0k^2(m^2-1),
\end{eqnarray*}
then we have the first type of soltion-cnoidal periodic wave interaction solution
\begin{eqnarray}\label{firstcaseeq}
\nonumber U &=& \frac{2\lambda}{\beta} + \frac{ b^2_0k^2}{2\beta} \left\{(m^2+3) -2[1-m{\rm sn}(\xi,m)]^2
+4m  {\rm cn}(\xi,m){\rm dn}(\xi,m)\tanh(\eta) \right.\\
&&\left. -2  [1+m{\rm sn}(\xi,m)]^2\tanh^2(\eta) \right\},
\end{eqnarray}
with
$\xi=b_0k(x-c_2g)$ and $\eta=\frac{1}{2}\{kg+\xi+\ln[{\rm dn}(\xi,m)-m{\rm cn}(\xi,m) ]\}$.

\begin{figure}[!htbp]
\centering
\includegraphics[width=4.77in,height=3.25in]{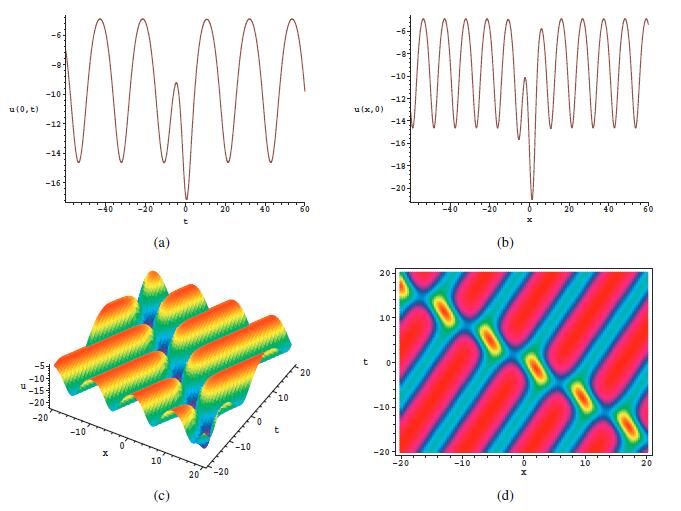}
\caption{(Color online) The first type of soliton-cnoidal wave interaction solution (\ref{firstcaseeq}) with the parameters $c_2=0.5,c_3=1,b_0=0.6,m=0.3,k=1$ and $g=t$: (a) the profile at $t = 0$; (b) the profile at $x = 0$; (c) the three-dimensional plot;  (d) the density plot.
\label{mix1d-fig1}}
\end{figure}

\begin{figure}[!htbp]
\centering
\includegraphics[width=4.77in,height=2.00in]{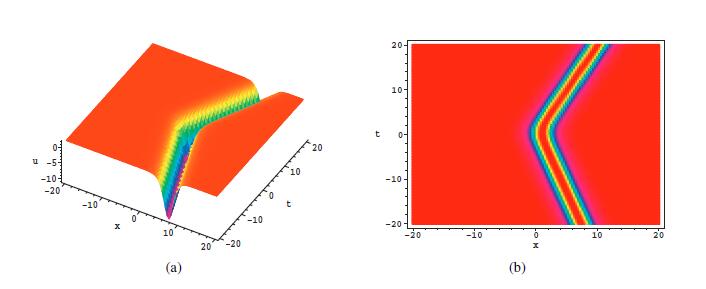}
\caption{(Color online) The first type of soliton-cnoidal wave interaction solution (\ref{firstcaseeq}) degenerates to the resonant two-soliton with the modulus $m=1$, and other parameters are the same as ones in Fig.1:  (a) the three-dimensional plot;  (b) the density plot.
\label{mix1d-fig2}}
\end{figure}

\begin{figure}[!htbp]
\centering
\includegraphics[width=4.77in,height=3.25in]{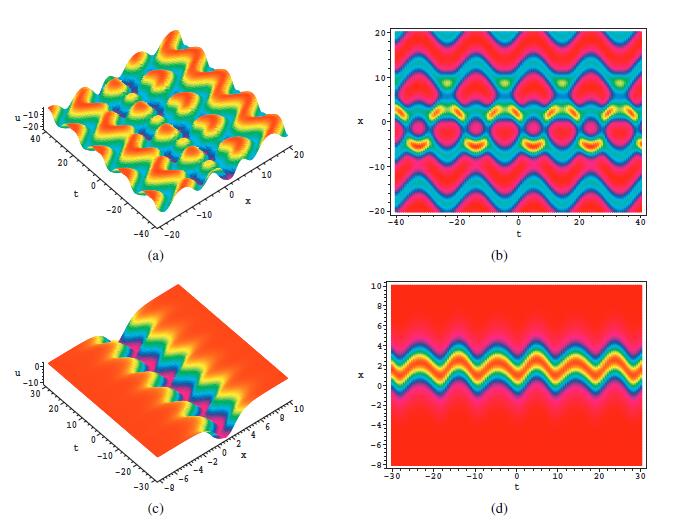}
\caption{(Color online) The first type of soliton-cnoidal wave interaction solution (\ref{firstcaseeq}) with the parameters $c_2=0.5,c_3=1,b_0=0.6,m=0.3,k=1$ and $g=5\sin(\frac{1}{3}t)$: (a) the three-dimensional plot; (b) the density plot. The degenerated case takes the same parameters except the modulus $m=1$: (c) the three-dimensional plot;  (d) the density plot.
\label{mix1d-fig3}}
\end{figure}

Two group of examples corresponding to the first type of soliton-cnoidal wave interaction solution (\ref{firstcaseeq}) under the different choices of the parameters are illustrated in Figs.\ref{mix1d-fig1}-\ref{mix1d-fig3}.
In Figs.\ref{mix1d-fig1} and \ref{mix1d-fig2}, we exhibit such interaction solution when the arbitrary function $g$ is chosen as a linear function $g=t$.
Clearly, one can observe that a dark soliton propagates on a cnoidal wave background instead of the plane continuous wave background in Fig.\ref{mix1d-fig1}.
In particular, when all other parameters are fixed but the the Jacobian elliptic function's modulus takes the limit $m=1$,
the periodic wave is reduced to solitary wave and the interaction solution becomes the resonant two-soliton, which is shown in Fig.\ref{mix1d-fig2}.
As another kind of illustrated case, we take the arbitrary function $g$ as a periodic function $g=5\sin(\frac{1}{3}t)$,
then the soliton-cnoidal wave interaction solution and the degenerated soliton move according to the periodic function $5\sin(\frac{1}{3}t)$ respectively, which are displayed distinctly in Fig.\ref{mix1d-fig3}.

\emph{Case 2} Another special solution of the reduction Eq. (\ref{last-reduce-1}) reads
\begin{eqnarray}
&& Q(z)=\frac{1}{b_0+b_1 {\rm sn}^2(dz,m)},
\end{eqnarray}
with
\begin{eqnarray*}
&& a_3=-\frac{k^2(3b^2_0m^2+2b_0b_1m^2+2b_0b_1+b^2_1)}{b_0(b_0+b_1)(b_0m^2+b_1)},\ \
 \beta=\frac{c_2}{2c_3}-\frac{c_2k^2(3b_0m^2+b_1m^2-3c_2m^2+b_1)}{2c_3b_0(b_0+b_1)(b_0m^2+b_1)},\\
&& \lambda=\frac{1}{4}+\frac{c_2k^2m^2}{4b_0(b_0+b_1)(b_0m^2+b_1)},\ \
 d=\frac{k}{2}\sqrt{\frac{-b_1}{b_0(b_0+b_1)(b_0m^2+b_1)}}.
\end{eqnarray*}
Thus, the second type of soltion-cnoidal periodic wave interaction solution has the form
\begin{eqnarray}\label{secondcaseeq}
\nonumber U&=&\frac{2\lambda}{\beta}+\frac{1}{\beta[b_0+b_1 {\rm sn}^2(\xi,m)]^2}
\bigg\{ \frac{k^2}{2} +2b_1d^2[b_0 - 2A_1{\rm sn}^2(\xi,m)
+ A_2{\rm sn}^4(\xi,m) ] \\
&&
-4 b_1 d k {\rm sn}(\xi,m){\rm cn}(\xi,m){\rm dn}(\xi,m)\tanh(\eta)
-k^2\tanh^2(\eta) \bigg\},
\end{eqnarray}
with $A_1=b_0+b_1+b_0m^2$, $A_2=3b_0m^2+b_1m^2+b_1$, $\xi=d(x-c_2g)$, $\eta=\frac{k}{2db_0}\{db_0 g +E_\pi[{\rm sn}(\xi,m), -\frac{b_1}{b_0},m]\}$
and $E_\pi$ is the third type of incomplete elliptic integral.

\begin{figure}[!htbp]
\centering
\includegraphics[width=4.77in,height=3.25in]{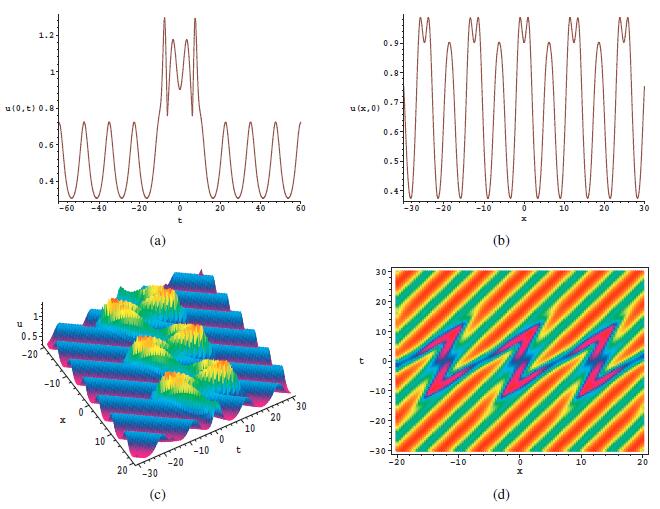}
\caption{(Color online) The second type of soliton-cnoidal wave interaction solution (\ref{secondcaseeq}) with the parameters $c_2=c_3=k=1,b_0=0.8,b_1=-0.3,m=0.9$ and $g=0.5t$: (a) the profile at $t=0$;
 (b) the profile at $x=0$; (c) the three-dimensional plot;  (d) the density plot .
\label{mix1d-fig4}}
\end{figure}
\begin{figure}[!htbp]
\centering
\includegraphics[width=4.77in,height=2.00in]{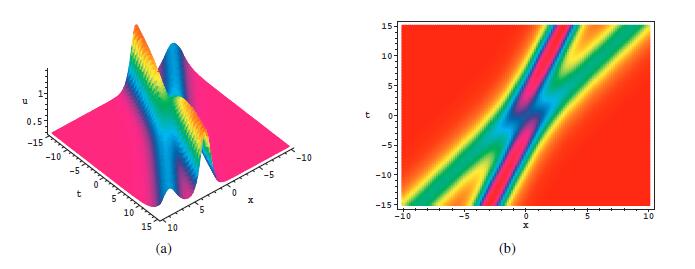}
\caption{(Color online) The second type of soliton-cnoidal waveinteraction solution (\ref{secondcaseeq}) degenerates to the two-soliton with the modulus $m=1$, and other parameters are the same as ones in Fig.\ref{mix1d-fig4}: (a) the three-dimensional plot;  (b) the density plot.
\label{mix1d-fig5}}
\end{figure}

\begin{figure}[!htbp]
\centering
\includegraphics[width=4.77in,height=3.25in]{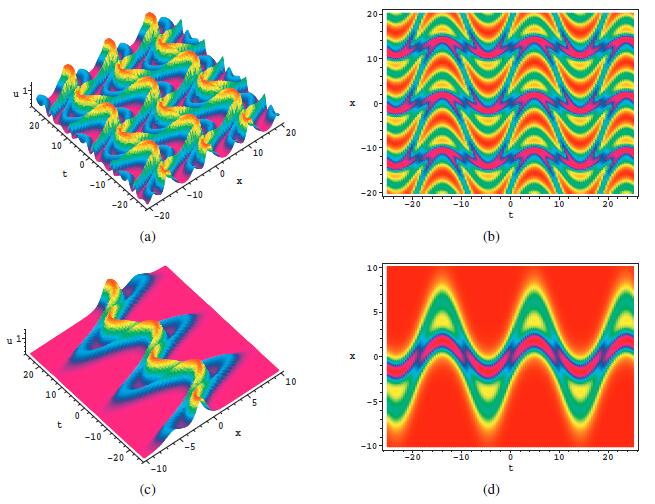}
\caption{(Color online)
 The second type of soliton-cnoidal wave interaction solution (\ref{secondcaseeq}) with the parameters $c_2=0.5,c_3=1,b_0=0.6,m=0.9,k=1$ and $g=5\sin(\frac{1}{3}t)$: (a) the three-dimensional plot; (b) the density plot. The degenerated case takes the same parameters except the
modulus $m=1$: (c) the three-dimensional plot;  (d) the density plot.
\label{mix1d-fig6}}
\end{figure}

For the second type of soliton-cnoidal wave interaction solution (\ref{firstcaseeq}), we still illustrate it graphically for two cases with the different values of parameters in Figs.\ref{mix1d-fig4}-\ref{mix1d-fig6}.
For the first case, the arbitrary function $g$ in (\ref{firstcaseeq}) is fixed as a linear function $g=0.5t$.
It can be seen that a bell-shaped bright soliton propagates on a cnoidal wave background in Fig.\ref{mix1d-fig4} when the modulus $m\neq1$,
while the degenerated case exhibits that two bright soliton undergo elastic collision when $m=1$ in Fig.\ref{mix1d-fig5}.
Similarly, if the arbitrary function $g$ is chosen as a periodic function $g=5\sin(\frac{1}{3}t)$,
then the soliton-cnoidal wave interaction solution and the two soliton move according to the taken periodic function.
These continuous nonlinear wave propagating behaviors are shown in Fig.\ref{mix1d-fig6} respectively.

\section{Summary and discussions}

In this paper, we study the nonlocal symmetry for the integrable SWW equation of AKNS type in classical shallow water wave theory.
Starting from Lax pair of the SWW-AKNS equation, nonlocal symmetry expressed by the square spectral function is derived.
Then infinitely many nonlocal symmetries are obtain by introducing the arbitrary spectrum parameter.
These nonlocal symmetries are localized by embodying some new auxiliary dependent variables and the SWW equation is extended to enlarged system.
By solving the initial value problem, Darboux transformation for the prolonged system is established due to the arbitrariness of the spectral parameter.
For the enlarged system, the standard Lie point symmetry is used to study similarity reductions and explicit group invariant solutions.
The results show that in these similar solutions, one can obtain a kind of exact soliton-cnoidal wave interaction solution, which represents soliton propagating on a cnoidal periodic wave background.
Interesting characteristics of the interaction solution between soliton and cnoidal periodic wave are displayed graphically and discussed analytically.
This type of solution is expected to applicable to the analysis of physically interesting processes in shallow water wave.

\section*{Acknowledgment}
This work was supported by the National Natural Science Foundation
 of China (Grant No. 11447017), the Natural Science Foundation of Zhejiang Province (Grant No.
 LY14A010005) and the Applied Nonlinear Science and Technology from the Most Important Among
 all the Top Priority Disciplines of Zhejiang Province.







\end{document}